\documentclass[aps,prl,letter,twocolumn,superscriptaddress,floatfix,nofootinbib,showpacs,amsmath,amssymb,altaffilletter]{revtex4}
\usepackage{graphicx}
\usepackage{dcolumn}
\usepackage{bm}
\usepackage{color}
\usepackage{verbatim}
\usepackage{paralist}
\usepackage{amssymb}
\usepackage[english]{babel}
\newcommand{\powert}[2]{\mbox{#2$\times$10$^{#1}$}}

\newcommand{\hydro}{\mbox{$^{1}$H}}
\newcommand{\ber}{\mbox{$^{7}$Be}}
\newcommand{\bele}{\mbox{$^{11}$Be}}
\newcommand{\btwe}{\mbox{$^{12}$B}}
\newcommand{\bis}{\mbox{$^{214}$Bi}}
\newcommand{\bor}{\mbox{$^{8}$B}}

\newcommand{\cfo}{\mbox{$^{14}$C}}
\newcommand{\ctwe}{\mbox{$^{12}$C}}
\newcommand{\cele}{\mbox{$^{11}$C}}
\newcommand{\cten}{\mbox{$^{10}$C}}

\newcommand{\radon}{\mbox{$^{222}$Rn}}
\newcommand{\thoron}{\mbox{$^{220}$Rn}}

\newcommand{\pofo}{\mbox{$^{214}$Po}}
\newcommand{\potwe}{\mbox{$^{212}$Po}}
\newcommand{\pote}{\mbox{$^{210}$Po}}
\newcommand{\bifo}{\mbox{$^{214}$Bi}}
\newcommand{\bitwe}{\mbox{$^{212}$Bi}}
\newcommand{\Bipo}{\mbox{$^{214}$Bi-$^{214}$Po}}
\newcommand{\bipo}{\mbox{$^{212}$Bi-$^{212}$Po}}
\newcommand{\tho}{\mbox{$^{232}$Th}}
\newcommand{\tal}{\mbox{$^{208}$Tl}}
\newcommand{\ura}{\mbox{$^{238}$U}}
\newcommand{\che}{\v{C}erenkov}
\newcommand{\Pee}{\mbox{$P_{ee}$}}
\newcommand{\oPee}{\mbox{$\overline{P}_{ee}$}}
\newcommand{\pep}{\mbox{\it pep}}
\newcommand{\pp}{\mbox{\it pp}}

\newcommand{\nue}{\mbox{$\nu_e$}}

\newcommand{\gfbx}{{\tt G4Bx}}

\begin{document}

\title{Measurement of the solar \bor~neutrino rate with a liquid scintillator target and  3\,MeV  energy threshold  in the Borexino detector}

\newcommand{\APC}{Laboratoire AstroParticule et Cosmologie, 75231 Paris cedex 13, France}
\newcommand{\Dubna}{Joint Institute for Nuclear Research, 141980 Dubna, Russia}
\newcommand{\Genova}{Dipartimento di Fisica, Universit\`a e INFN, Genova 16146, Italy}
\newcommand{\Heidelberg}{Max-Planck-Institut f\"ur Kernphysik, 69029 Heidelberg, Germany}
\newcommand{\Kiev}{Kiev Institute for Nuclear Research, 06380 Kiev, Ukraine}
\newcommand{\Krakow}{M.~Smoluchowski Institute of Physics, Jagiellonian University, 30059 Krakow, Poland}
\newcommand{\Kurchatov}{RRC Kurchatov Institute, 123182 Moscow, Russia}
\newcommand{\LNGS}{INFN Laboratori Nazionali del Gran Sasso, SS 17 bis Km 18+910, 67010 Assergi (AQ), Italy}
\newcommand{\Milano}{Dipartimento di Fisica, Universit\`a degli Studi e INFN, 20133 Milano, Italy}
\newcommand{\Moscow}{Institute of Nuclear Physics, Lomonosov Moscow State University, 119899, Moscow, Russia}
\newcommand{\Munich}{Physik Department, Technische Universit\"at Muenchen, 85747 Garching, Germany}
\newcommand{\Perugia}{Dipartimento di Chimica, Universit\`a e INFN, 06123 Perugia, Italy}
\newcommand{\Peters}{St. Petersburg Nuclear Physics Institute, 188350 Gatchina, Russia}
\newcommand{\Princeton}{Physics Department, Princeton University, Princeton, NJ 08544, USA}
\newcommand{\PrincetonChemEng}{Chemical Engineering Department, Princeton University, Princeton, NJ 08544, USA}
\newcommand{\UMass}{Physics Department, University of Massachusetts, Amherst, MA 01003, USA}
\newcommand{\Virginia}{Physics Department, Virginia Polytechnic Institute and State University, Blacksburg, VA 24061, USA}

\author{G.~Bellini}\affiliation{\Milano}
\author{J.~Benziger}\affiliation{\PrincetonChemEng}
\author{S.~Bonetti}\affiliation{\Milano}
\author{M.~Buizza~Avanzini}\affiliation{\Milano}
\author{B.~Caccianiga}\affiliation{\Milano}
\author{L.~Cadonati}\affiliation{\UMass}
\author{F.~Calaprice}\affiliation{\Princeton}
\author{C.~Carraro}\affiliation{\Genova}
\author{A.~Chavarria}\affiliation{\Princeton}
\author{A.~Chepurnov}\affiliation{\Moscow}
\author{F.~Dalnoki-Veress}\affiliation{\Princeton}
\author{D.~D'Angelo}\affiliation{\Milano}
\author{S.~Davini}\affiliation{\Genova}
\author{H.~de~Kerret}\affiliation{\APC}
\author{A.~Derbin}\affiliation{\Peters}
\author{A.~Etenko}\affiliation{\Kurchatov}
\author{K.~Fomenko}\affiliation{\Dubna}
\author{D.~Franco}\affiliation{\Milano}
\author{C.~Galbiati}\affiliation{\Princeton}
\author{S.~Gazzana}\affiliation{\LNGS}
\author{C.~Ghiano}\affiliation{\LNGS}
\author{M.~Giammarchi}\affiliation{\Milano}
\author{M.~Goeger-Neff}\affiliation{\Munich}
\author{A.~Goretti}\affiliation{\Princeton}
\author{E.~Guardincerri}\affiliation{\Genova}
\author{S.~Hardy}\affiliation{\Virginia}
\author{Aldo~Ianni}\affiliation{\LNGS}
\author{Andrea~Ianni}\affiliation{\Princeton}
\author{M.~Joyce}\affiliation{\Virginia}
\author{G.~Korga}\affiliation{\LNGS}
\author{D.~Kryn}\affiliation{\APC}
\author{M.~Laubenstein}\affiliation{\LNGS}
\author{M.~Leung}\affiliation{\Princeton}
\author{T.~Lewke}\affiliation{\Munich}
\author{E.~Litvinovich}\affiliation{\Kurchatov}
\author{B.~Loer}\affiliation{\Princeton}
\author{P.~Lombardi}\affiliation{\Milano}
\author{L.~Ludhova}\affiliation{\Milano}
\author{I.~Machulin}\affiliation{\Kurchatov}
\author{S.~Manecki}\affiliation{\Virginia}
\author{W.~Maneschg}\affiliation{\Heidelberg}
\author{G.~Manuzio}\affiliation{\Genova}
\author{Q.~Meindl}\affiliation{\Munich}
\author{E.~Meroni}\affiliation{\Milano}
\author{L.~Miramonti}\affiliation{\Milano}
\author{M.~Misiaszek}\affiliation{\Krakow}\affiliation{\LNGS}
\author{D.~Montanari}\affiliation{\LNGS}\affiliation{\Princeton}
\author{V.~Muratova}\affiliation{\Peters}
\author{L.~Oberauer}\affiliation{\Munich}
\author{M.~Obolensky}\affiliation{\APC}
\author{F.~Ortica}\affiliation{\Perugia}
\author{M.~Pallavicini}\affiliation{\Genova}
\author{L.~Papp}\affiliation{\LNGS}
\author{L.~Perasso}\affiliation{\Milano}
\author{S.~Perasso}\affiliation{\Genova}
\author{A.~Pocar}\affiliation{\UMass}
\author{R.S.~Raghavan}\affiliation{\Virginia}
\author{G.~Ranucci}\affiliation{\Milano}
\author{A.~Razeto}\affiliation{\LNGS}
\author{A.~Re}\affiliation{\Milano}
\author{P.~Risso}\affiliation{\Genova}
\author{A.~Romani}\affiliation{\Perugia}
\author{D.~Rountree}\affiliation{\Virginia}
\author{A.~Sabelnikov}\affiliation{\Kurchatov}
\author{R.~Saldanha}\affiliation{\Princeton}
\author{C.~Salvo}\affiliation{\Genova}
\author{S.~Sch\"onert}\affiliation{\Heidelberg}
\author{H.~Simgen}\affiliation{\Heidelberg}
\author{M.~Skorokhvatov}\affiliation{\Kurchatov}
\author{O.~Smirnov}\affiliation{\Dubna}
\author{A.~Sotnikov}\affiliation{\Dubna}
\author{S.~Sukhotin}\affiliation{\Kurchatov}
\author{Y.~Suvorov}\affiliation{\Milano}\affiliation{\Kurchatov}
\author{R.~Tartaglia}\affiliation{\LNGS}
\author{G.~Testera}\affiliation{\Genova}
\author{D.~Vignaud}\affiliation{\APC}
\author{R.B.~Vogelaar}\affiliation{\Virginia}
\author{F.~von~Feilitzsch}\affiliation{\Munich}
\author{J.~Winter}\affiliation{\Munich}
\author{M.~Wojcik}\affiliation{\Krakow}
\author{A.~Wright}\affiliation{\Princeton}
\author{M.~Wurm}\affiliation{\Munich}
\author{J.~Xu}\affiliation{\Princeton}
\author{O.~Zaimidoroga}\affiliation{\Dubna}
\author{S.~Zavatarelli}\affiliation{\Genova}
\author{G.~Zuzel}\affiliation{\Heidelberg}

\collaboration{Borexino Collaboration}

\date{\today}

\begin{abstract}

We report the measurement of $\nu$-$e$ elastic scattering  from \bor\  solar neutrinos with 3\,MeV energy threshold by the Borexino detector in Gran Sasso (Italy).  The rate of solar neutrino-induced electron scattering events above this energy in Borexino is $0.217\pm 0.038 (stat)\pm 0.008 (syst)$~cpd/100\,t, which corresponds to $\Phi^{\rm ES}_{\rm ^8B}$ = {2.4 $\pm$ 0.4$\pm$ 0.1}$\times$10$^6$~cm$^{-2}$ s$^{-1}$, in good agreement with measurements from SNO and SuperKamiokaNDE.  Assuming the \bor\ neutrino flux predicted by the high metallicity Standard Solar Model, the average \bor\ \nue\ survival probability above 3 MeV is measured to be 0.29$\pm$0.10.  The survival probabilities for \ber\ and \bor\ neutrinos as measured by Borexino differ by  1.9 $\sigma$.  These results are consistent with the prediction of the MSW-LMA solution of a transition in the solar \nue\ survival probability \Pee\ between the low energy vacuum-driven and the high-energy matter-enhanced solar neutrino oscillation regimes.

\end{abstract}




\maketitle

\section{Introduction}
\label{sec:intro}

Solar \bor-neutrino  spectroscopy  has been so far performed by the water \che\ detectors KamiokaNDE, SuperKamiokaNDE, and SNO~\cite{Hir89,SKII08,SKI05,SNO07}.  The first two experiments used elastic $\nu$-$e$ scattering for the detection of neutrinos, whereas SNO also exploited nuclear reaction channels on deuterium with heavy water as  target.  These experiments provided robust spectral measurements with $\sim$5\,MeV threshold or higher for scattered electrons; a recent SNO analysis reached a 3.5\,MeV threshold~\cite{SNO09}.  

We report the first observation of solar \bor-neutrinos with a liquid scintillator detector, performed by the Borexino experiment~\cite{BXD08,BX09} via elastic $\nu$-$e$ scattering.  Borexino is the first experiment to succeed in suppressing all
  major backgrounds, above the 2.614\,MeV $\gamma$ from the decay of \tal,  to a rate below that of electron scatterings from solar neutrinos.  This allows to reduce  the energy threshold for scattered electrons  by \bor\ solar neutrinos to 3\,MeV, the lowest ever reported for the electron scattering channel. To facilitate a comparison to the results of SuperKamiokaNDE \cite{SKI05} and SNO  D$_2$O phase \cite{SNO07}, we also report the measured \bor\ neutrino interaction rate with 5\,MeV threshold. 

Since Borexino also detected low energy solar \ber\ neutrinos~\cite{BX07,BX08}, this is the first experiment where both branches of the solar \pp-cycle have been measured simultaneously  in the same target.  The large mixing angle solution (LMA) of the MSW effect~\cite{MSW} predicts a transition in the $\nu_e$ survival probability from the vacuum oscillation regime at low energies to the matter dominated regime at high energies.  Results on solar \ber\ and \bor\ neutrinos from Borexino, combined with prediction on the absolute neutrino fluxes from the Standard Solar Model~\cite{BS07,Pen08,Ser09}, confirm that our data are in agreement with the MSW-LMA prediction within 1$\sigma$.

\section{Experimental Apparatus and Energy Threshold}
\label{sec:detector}

The Borexino detector is located at the underground Laboratori Nazionali del Gran Sasso (LNGS) in central Italy, at a depth of 3600\,m.w.e..  Solar neutrinos are detected in Borexino exclusively via elastic $\nu$-$e$ scattering in a liquid scintillator.  The active target consists of 278\,t of pseudocumene (PC, 1,2,4-trimethylbenzene), doped with 1.5\,g/l of PPO (2,5-diphenyloxazole, a fluorescent dye).  The scintillator is contained in a thin (125\,$\mu$m) nylon vessel of 4.25\,m nominal radius, and is shielded by two concentric PC buffers (323 and 567\,t) doped with 5.0\,g/l of a scintillation light quencher (dimethylphthalate). Scintillator and buffers are contained in a Stainless Steel Sphere (SSS) with a diameter of 13.7\,m and the scintillation light is detected via 2212 8''~photomultiplier tubes (PMTs) uniformly distributed on the inner surface of the SSS. The two PC buffers are separated by a second thin nylon membrane to prevent diffusion of  the radon emanated  by the PMTs and by the stainless steel of the sphere into the scintillator. The SSS is enclosed in a 18.0 m diameter, 16.9 m high domed Water Tank (WT), containing 2100\,t of ultra-pure water as an additional shielding against external gamma- and neutron background. 208 8''~PMTs in the WT detect the \che\ light produced by muons in the water shield, serving as a highly efficient muon veto.  A complete description of the Borexino detector can be found in Ref.~\cite{BXD08}.

Scintillator detectors, with their high light yield, are sensitive to lower energy events than \che\ detectors.  In this analysis, the 3\,MeV energy threshold is imposed mainly by the 2.614\,MeV $\gamma$-rays from the decay of \tal\ (\tho\ chain, $Q$=5.001\,MeV) in the PMTs and in the SSS and by the finite energy resolution of the detector: a tail of 2.6 MeV $\gamma$ events leaks at higher energies, along with a very small percentage of combined $\gamma$'s from \bis\ (\ura\ chain, $Q$=3.272\,MeV). The  3\,MeV energy threshold eliminates sample contamination from such events.

Potential background sources above 3\,MeV include the radioactive decays of residual \bis\  and \tal\  within the liquid scintillator, decays of cosmogenic isotopes (see Table~\ref{tab:cosmogenic} later), high energy $\gamma$-rays from neutron capture, and cosmic muons.  No background from $\alpha$ decays is expected at these energies, since the light quenching of $\alpha$'s in organic liquid scintillators reduces their visible energy in the electron-equivalent scale below 1\,MeV.  A measurement of \bor\ neutrinos with a 3\,MeV energy threshold is contingent upon high radiopurity of the scintillator target.  \ura\ and \tho\ concentrations in the Borexino scintillator have been measured at (1.6$\pm$0.1$)\times$10$^{-17}$\,g/g and (6.8$\pm$1.5)$\times$10$^{-18}$\,g/g, respectively, and record low levels of backgrounds in the energy range 0.2-5.0\,MeV have been reported~\cite{BXD08,BX08}.  The dominant background in the energy range of interest for solar \bor~neutrinos originates from spallation processes of high energy cosmic muons.  This paper demonstrates that, thanks to the LNGS depth and the Borexino muon veto system, cosmogenic background can be reduced below the rate of interaction of \bor\ neutrinos, thus allowing the neutrino rate to be measured.

\section{Energy and Position Response of the Detector and Associated Systematics}
\label{sec:calibration}

We tuned the response of  energy and position reconstruction algorithms with a dedicated calibration campaign.  We used an off-axis source insertion  system designed to position radioactive and/or luminous sources at several locations throughout the detector active target. The source position can be measured with a set of stereoscopic cameras installed on the SSS with an uncertainty of $\pm$2\,cm in $x$, $y$, and $z$ \cite{Bac04} ($\pm$3.5\,cm in radius).

As explained above, Borexino has a unique sensitivity to electron scattering from low energy solar neutrinos, thanks to its unmatched record on background below the natural radioactivity barrier.  To preserve this capability, the off-axis source insertion  system was designed to respect stringent limits on leak tightness and cleanliness of the mechanics in contact with the liquid scintillator.  A detailed description of the technique used for the calibration and for the position reconstruction in Borexino is in preparation.

\subsection{Energy Scale}\label{sec:energy}

Calibration of the energy scale allows to establish with high confidence the energy threshold for the \bor~neutrino analysis and the error in its determination, and to calibrate the energy scale to allow the energy spectrum of the electrons scattered by \bor\ neutrinos to be determined.  Distortions of the energy scale are due to physical  effects  (quenching), geometrical  effects (light collection), and to the electronics, which was designed for optimal performance in the low-energy range of \ber\ neutrinos,  in a regime where a single photoelectron is expected for each PMT. 
At higher energies, the electronics response to  multiple photoelectron hits on a single channel is not linear.  
For each triggered channel, the charge from photoelectrons in a 80\,ns gate is integrated and recorded, but photoelectrons in the following 65\,ns dead time window are lost.
The resulting fraction of  lost charge  increases with energy and can reach  $\sim$10\% at $\sim$10 MeV.  Moreover, the number of detected photoelectrons   depends on the event position in the active volume, due to  differences in PMT coverage.  For an accurate determination of the energy scale, the dominant non-linearities  have been reproduced with a  Monte Carlo simulation  and with calibration measurements.

To avoid contamination during calibration, $\beta$~sources were not put in direct contact with the scintillator. Instead, we calibrated the response of the detector with encapsulated $\gamma$~sources. The scintillation induced by $\gamma$-rays is due to the ionizing tracks of the secondary electrons, thus, the two energy scales are closely related.

Establishing a correlation between the $\beta$ and $\gamma$ energy scales required extensive simulations with the \gfbx\ Monte Carlo code.  \gfbx\ is based on {\tt Geant4}~\cite{Ago03,All06} and simulates in detail all of the detector component, and includes scintillation,  \che\ photon production,  absorption and scattering of light in the scintillator and in the buffer, as well as the PMT response.  Each secondary electron in a $\gamma$-induced Compton electron cascade is affected by  energy-dependent ionization quenching, which amplifies the distortion in the $\gamma$ energy scale. 
The quenching effect is modeled with the Birks formalism~\cite{Bir51}.  
A second package, {\tt BxElec}, simulates in detail the response of the electronics.
Finally, Monte Carlo data are processed by the same reconstruction code used for real data. A detailed description of the Monte Carlo codes and of the energy reconstruction algorithm is  in preparation.  

\begin{figure}
\begin{center}
\includegraphics[width=0.49\textwidth]{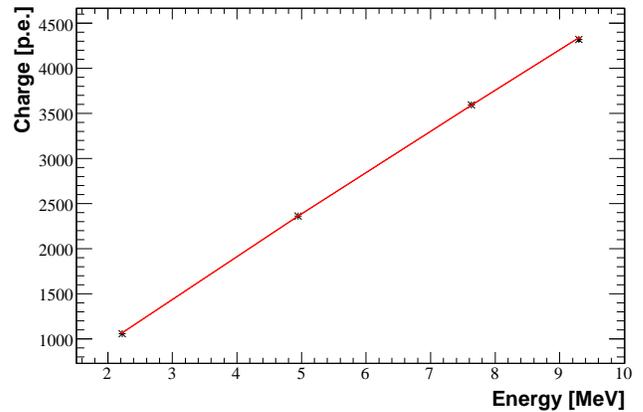}
\caption{ Black dots are the measured peak positions of $\gamma$ radiation  induced by neutron captures in $^{1}$H (2.223 MeV), $^{12}$C (4.945 MeV), $^{56}$Fe (7.631 MeV) and $^{54}$Fe (9.298 MeV) in the detector center. Red line is the Monte Carlo prediction for $\gamma$ rays generated  in the detector center.}
\label{fig:enecal}
\end{center}
\end{figure}

To calibrate the detector energy response to   \bor\ neutrinos, we used an $^{241}$Am$^9$Be neutron source positioned at the center of the detector and at several positions at 3~m radius.
Neutron capture  on $^1$H and on $^{12}$C in the scintillator results in the emission of $\gamma$-rays from the 2.223\,MeV and 4.945\,MeV excited states, respectively. In addition, neutron capture on the stainless steel  of the insertion system  produces $\gamma$-rays from the 7.631\,MeV ($^{56}$Fe) and 9.298\,MeV ($^{54}$Fe) excited states. We validate the Monte Carlo code by simulating  the four $\gamma$-rays in both the positions. In Figure~\ref{fig:enecal} we show the results of the calibration of the $\gamma$-equivalent energy scale in the detector center.  Monte Carlo  simulations reproduce $\gamma$ peak positions and resolutions at  $\sigma_1$ = 1\%  precision in the detector center (as shown in Figure \ref{fig:enecal}), and  at $\sigma_2$ = 4\% precision at 3\,m from the detector's center. Assuming the same accuracy for the $\beta$-equivalent energy scale, we extrapolate it by simulating electrons uniformly distributed in the scintillator, and then selecting those with  reconstructed position within the fiducial volume. The error on the energy scale is obtained with a linear interpolation from $\sigma_1$ in the detector center to $\sigma_2$ at 3 m, along the radius.  The $\beta$-equivalent energy scale, in the energy region above 2~MeV,  can  be parametrized as:

\begin{equation}
N = a \cdot E + b,
\label{eq1}
\end{equation}

\noindent where $N$ is the number of photoelectrons (p.e.) detected by the PMTs, $a$=459$\pm$11\,p.e./MeV and $b$=115$\pm$38\,p.e..  The  non-zero intercept $b$ is related to the fact that this  description is valid only in this energy range and that the overall relation between $N$ and $E$ is non linear.

The anticipated 3\,MeV (5\,MeV) energy threshold for the \bor\ analysis corresponds to 1494\,p.e. (2413\,p.e.) within a 3\,m radial distance from the center of the detector.  The uncertainty associated to the 3\,MeV (5\,MeV) energy threshold is obtained by propagating the errors of Eq.~\ref{eq1} and is equal to 51 p.e. (68 p.e.).

\subsection{Vertex Reconstruction}
The positions of scintillation events are reconstructed  with a photon time-of-flight method.  We computed with \gfbx\ a probability density function (PDF) for the time of transit of photons from their emission point to their detection as photoelectron signals in the electronics chain.  We refined the PDF with data collected in the calibration campaign. Event coordinates ($x_0$, $y_0$, $z_0$) and time ($t_0$) are obtained by minimizing:

\begin{equation}
{\cal{L}}(x_0,y_0,z_0,t_0) = \prod_i {\rm PDF}\left( t_i - t_0 - \frac{d_{0,i} \cdot n_{\rm eff}}{c} \right)
\end{equation}

\noindent where the index $i$ runs over the triggered PMTs,  $t_i$ is the time of arrival of the photoelectron on the $i^{\rm th}$ electronic channel, and $d_{0,i}$ is the distance from the event position and the $i^{\rm th}$ PMT. $n_{\rm eff}$ is an empirically-determined effective index of refraction 
to account for any other effect that is not accounted for in the reconstruction algorithm but impacts the distribution of PMT hit times, both in the optics (e.g. Rayleigh scattering) and the electronics (e.g. multiple photoelectron occupancy).

\begin{figure}[!t]
\begin{center}
\includegraphics[width=0.49\textwidth]{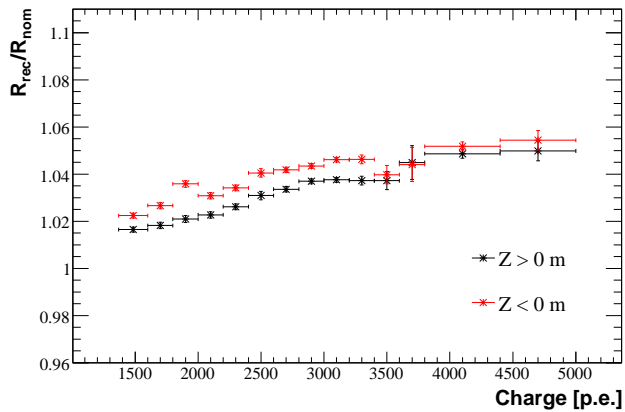}
\caption{Ratio of the reconstructed radial position of $\gamma$ events from the $^{241}$Am$^9$Be source in Borexino to the source radial position measured by the CCD camera system, as a function of the measured charge.}
\label{fig:recon}
\end{center}
\end{figure}

The Borexino electronics records the time of each detected photoelectron introducing a dead time of 145 ns after each hit for each individual channel.  Therefore, the timing distribution is biased at high energy, where multiple photoelectrons are detected by each channel, and the position reconstruction is energy dependent.
To measure this effect, we deployed the $^{241}$Am$^9$Be  neutron source  at the six cardinal points of the sphere defining the fiducial volume, {\it i.e.} those points lying on axis through the center of the detector, with off-center coordinates from the set $x$=$\pm$3\,m, $y$=$\pm$3\,m, and $z$=$\pm$3\,m. The recoiled proton from neutron scattering allows us to study the reconstructed position as function of the collected charge up to $\sim$5000 p.e.. Figure~\ref{fig:recon} shows the ratio of measured versus nominal position of the $^{241}$Am$^9$Be source. This data was used to define  the fiducial volume $R_{\rm nom}$$<$3\,m.
The non-homogeneous distribution of live PMTs, in particular the large deficit of live PMTs in the bottom hemisphere \cite{BXD08},  is responsible for the different spatial response at mirrored positions about the $x$-$y$ plane.  Thus, as shown in  Figure~\ref{fig:recon}, two radial functions have been defined for positive and negative $z$ positions.

After all post-calibration improvements to the event reconstruction algorithm, typical resolution in the event position reconstruction is 13$\pm$2\,cm in $x$ and $y$, and 14$\pm$2\,cm in $z$ at the relatively high \bifo\ energies.  The spatial resolution is expected to scale as $1/\sqrt{N}$ where $N$ is the number of triggered PMTs, and this was confirmed by determining the \cfo\ spatial resolution to be 41$\pm$6\,cm (1$\sigma$) at 140\,keV~\cite{BXD08,BX08}.  Systematic deviations of reconstructed positions from the nominal source position are due to the 3.5\,cm accuracy of the CCD cameras in the determination of the calibration source position, and in 1.6 cm introduced by the energy dependency. The overall systematics are within 3.8\,cm throughout the 3\,m-radius fiducial volume.

\section{\bor-neutrino flux}\label{sec:b8nu}

We report our results for the rate of electron scattering above 3\,MeV from \bor\ neutrino interactions in the active target.  We also report the result above the threshold of 5\,MeV, to facilitate the comparison with results reported by SNO~\cite{SNO07} and SuperKamiokaNDE phase-I~\cite{SKI05} at the same threshold.  This energy range is unaffected by the scintillator intrinsic background, since the light quenching effect reduces the visible energy of \tal\ ($Q$=5.001\,MeV) from $^{232}$Th contamination in the scintillator below the energy threshold of 5\,MeV.

The analysis in this paper is based on 488~live days of data acquisition, between July 15, 2007 and August 23, 2009, with a target mass of 100\,t, defined by a fiducial volume cut of radius 3\,m. The total exposure, after applying all the analysis cuts listed in the next section, is 345.3 days.

\subsection{Muon Rejection}

The cosmic muon rate at LNGS is 1.16$\pm$0.03\,m$^{-2}$hr$^{-1}$ with an average energy of 320$\pm4_{\rm stat}\pm11_{\rm sys}$\,GeV~\cite{MAC99}.  Each day, $\sim$4300~muons deposit energy in Borexino's inner detector.  
Depending on deposited energy and track length, there is a small but non-zero chance that a cosmic muon induces a number of photoelectrons comparable to the multi-MeV electron scatterings of interest for this analysis, and is mistaken for a point-like scintillation event.  A measurement of the neutrino interaction rate in Borexino requires high performance rejection of muon events and an accurate estimate of the muon tagging efficiency.

As  mentioned earlier, the Borexino WT is instrumented with 208 PMTs to serve as a muon veto.  If an Inner Detector (ID) event coincides in time with an Outer Detector (OD) trigger (i.e. more than 6~PMTs in the WT are hit within  a 150\,ns window),  the event is tagged as  muon and rejected.  However, the OD efficiency is not unity and  depends on the  direction of the incoming cosmic muon.  

In addition, we perform pulse-shape discrimination on the  hit time distribution of inner detector PMTs, since  for track-like events, like muons, such distribution generally extends to longer times than for point-like events, like $\beta$-decays and $\nu-e$ scattering.
We exclude muons from the event sample in the energy range of interest (3.0--16.3\,MeV, or 1413--6743\,p.e.)  by imposing the following requirements (ID cuts):

\begin{compactitem}
\item The  peak of the  reconstructed hits time distribution, with respect to the first hit, is between 0\,ns and 30\,ns.
\item The mean value of the  reconstructed hits time distribution, with respect to the first hit, is between 0\,ns and 100\,ns.
\end{compactitem}


The efficiency of the selection cuts was evaluated on a sample of 2,170,207~events, identified by the OD as muons. 
Only 22 of these events, a fraction of (1.0$\pm$0.2)$\times10^{-5}$, survive the ID cuts in the energy and spatial region of interest, and are tagged as possible scintillation events.

We do not have an absolute value for the OD muon veto efficiency, but we estimate it to be larger than 99\%,  from  \gfbx\ simulations. The residual muon rate, due to the combined inefficiency of the two tagging systems, taking into account the fact that the two detectors are independent, is \powert{-4}{\mbox{(4.5$\pm$0.9)}}\,muons/day/100\,t, or \powert{-4}{\mbox{(3.5$\pm$0.8)}}\,muons/day/100\,t above 5\,MeV.

\subsection{Cosmogenic Background Rejection}

\begin{table*}
\centering
\begin{tabular}{lccccclll}
\hline\hline
Isotopes		&$\tau$	 &$Q$	&Decay  &Expected Rate   & Fraction    &Expected Rate $>3~MeV$ &Measured Rate $>3~MeV$\\
			&	 &[MeV] &	&[cpd/100\,t] & $>3~MeV$ &   [cpd/100\,t]&  [cpd/100\,t]\\
\hline
$^{12}$B    &0.03\,s&  13.4   &$\beta^-$ & 1.41 $\pm$ 0.04 & 0.886 &1.25$\pm$ 0.03 & 1.48 $\pm$ 0.06\\
\hline
$^{8}$He    &0.17\,s&  10.6   &$\beta^-$ & 0.026 $\pm$ 0.012 & 0.898 & &\\
$^{9}$C	    &0.19\,s&  16.5   &$\beta^+$ & 0.096 $\pm$ 0.031 & 0.965 & (1.8 $\pm$0.3 )$\times$10$^{-1}$&(1.7 $\pm$ 0.5)$\times$10$^{-1}$\\
$^{9}$Li    &0.26\,s&  13.6   &$\beta^-$ & 0.071 $\pm$ 0.005 & 0.932& &\\
\hline
$^{8}$B	    &1.11\,s&  18.0   &$\beta^+$ & 0.273 $\pm$ 0.062 & 0.938& \\  	 
$^{6}$He    &1.17\,s&  3.5    &$\beta^-$ & NA & 0.009 &(6.0 $\pm$  0.8)$\times$10$^{-1}$ & (5.1 $\pm$  0.7)$\times$10$^{-1}$\\ 
$^{8}$Li    &1.21\,s&  16.0   &$\beta^-$ & 0.40 $\pm$ 0.07 & 0.875& & \\
\hline
$^{10}$C    &27.8\,s&  3.6    &$\beta^+$ & 0.54 $\pm$ 0.04 & 0.012 & (6.5$\pm$0.5)$\times$10$^{-3}$& (6.6$\pm$1.8) $\times$10$^{-3}$\\ 	 
\hline
$^{11}$Be   &19.9\,s&  11.5   &$\beta^-$ & 0.035 $\pm$ 0.006 & 0.902 & (3.2 $\pm$ 0.5)$\times$10$^{-2}$& (3.6$\pm$3.5)$\times$10$^{-2}$ \\
\hline\hline
\end{tabular}
\vspace{2mm}
\caption{Expected muon-induced contaminants with Q-value $>$ 3~MeV in Borexino.  The expected rate is obtained by extrapolating the recently  released data by the KamLAND Collaboration~\cite{Abe09} in accordance with the empirical law defined in Eq \ref{eq:scaling}.}
\label{tab:cosmogenic}
\end{table*}

\subsubsection{Fast cosmogenic veto}\label{sec:fast}

\begin{figure}[]
\begin{center}
\includegraphics[width=0.49\textwidth]{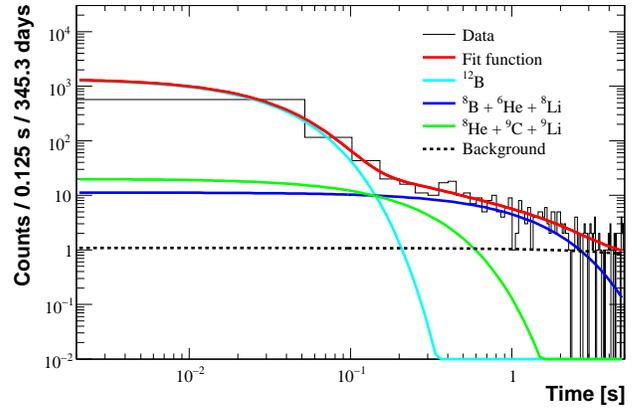}
\caption{ Cumulative  distribution of events with energy $>$ 3 MeV within a 5~s window after a muon in Borexino. The time distribution has been fit to three decay exponentials. The ensuing exponential lifetimes are $\tau =$ 0.031$\pm$0.002\,s, 0.25$\pm$0.21\,s, 1.01$\pm$0.36\,s and corresponds to the contribution from \btwe, $^{8}$He $+$ $^{9}$C $+$ $^{9}$Li and $^{8}$B $+$ $^{6}$He $+$ $^{8}$Li, respectively. The expected and measured rates for these cosmogenic isotopes are summarized in Table~\ref{tab:cosmogenic}.}
\label{fig:TimeAfterMuon}
\end{center}
\end{figure}

Table~\ref{tab:cosmogenic} presents a list of  expected cosmogenic isotopes produced by muons in Borexino.
The short-lived cosmogenics ($\tau < 2s$), as well as the $\gamma$-ray capture on \ctwe, are rejected by a 6.5\,s cut after each muon, with a  29.2\% fractional dead time.  Figure~\ref{fig:TimeAfterMuon} shows the time distribution of events following a muon.  The data is well fit by three exponentials with characteristic times of 0.031$\pm$0.002\,s (\btwe), 0.25$\pm$0.21\,s ($^{8}$He, $^{9}$C, $^{9}$Li), 1.01$\pm$0.36\,s ($^{8}$B, $^{6}$He , $^{8}$Li), in good agreement with the lifetimes of the short-lived isotopes (see Table~\ref{tab:cosmogenic}). From the fit we  estimate  the production rates of these cosmogenic isotopes in Borexino. We conclude that rejection of events in a 6.5\,s window following every muon crossing the SSS reduces the residual contamination of the short lived isotopes to \powert{-3}{\mbox{(1.7$\pm$0.2)}}\,cpd/100\,t (\powert{-3}{\mbox{(1.3$\pm$0.2)}}\,cpd/100\,t above 5~MeV).

The expected rates (R) quoted in Table~\ref{tab:cosmogenic} are obtained by scaling the production rates (R$^0$) measured by KamLAND~\cite{Abe09} with:
\begin{equation}
R = R^0\left(\frac{E_{\mu}}{E_{\mu}^0}\right)^\alpha \frac{\Phi_{\mu}}{\Phi_{\mu}^0} ,
\label{eq:scaling}\end{equation}
where  E$_{\mu}$ and $\Phi_{\mu}$ are the Borexino mean muon energy (320$\pm4_{\rm stat}\pm11_{\rm sys}$\,GeV) and flux (1.16$\pm$0.03\,m$^{-2}$ hr$^{-1}$), as measured by MACRO \cite{MAC99}, and E$_{\mu}^0$ (260$\pm$4\,GeV) and $\Phi_{\mu}^0$  (5.37$\pm$0.41 \,m$^{-2}$ hr$^{-1}$) are the corresponding KamLAND values.  
$\alpha$ is a scaling parameter  to relate cosmogenic production rate at different mean energies of the incoming muon flux; it is obtained in Ref.~\cite{Abe09} by fitting the production yield  of each isotope, simulated by FLUKA, as a function of muon beam energy.  Overall, Borexino data results are in agreement with the values quoted in Table~\ref{tab:cosmogenic}  within 15\%.

\subsubsection{Neutron rejection}\label{sec:neutron}

The cosmogenic background in Borexino includes  decays of radioactive isotopes due to spallation processes on the \ctwe\ nuclei in the scintillator, as well as the $\gamma$-rays from the capture of neutrons that are common by-products of such processes.  The capture time for neutrons in the Borexino scintillator has been measured to be 256.0$\pm$0.4\,$\mu$s, using a neutron calibration source, and the energy of the dominant $\gamma$-rays from neutron capture on \hydro\ at 2.223\,MeV is below the energy threshold of the present analysis.  On the other hand, the 4.9\,MeV $\gamma$-rays from neutron captures on $\ctwe$ is a potential background for this analysis. The rate is estimated by scaling the cosmogenic neutron capture rate on $^1$H by the fraction of captures on $\ctwe$ with respect to the total, measured with the $^{241}$Am$^9$Be neutron source. The neutron capture rate on $\ctwe$  is 0.86$\pm$0.01\,cpd/100 t. 

The fast cosmogenic veto, described in the {\it Fast Cosmogenic Veto} section, rejects neutrons produced in the scintillator or in the buffer by muon spallation with 99.99\% efficiency. To reject  neutrons produced in water, a second 2~ms veto is applied after each muon crossing the Water Tank only. The rejection efficiency for neutrons produced in water is 0.9996. The overall survival neutron rate in the energy range of interest and in the fiducial volume is \powert{-5}{(8.6$\pm$0.1)}\,cpd/100~t.

\subsubsection{\cten\ identification and subtraction}

A separate treatment is required for long-lived ($\tau$$>$2\,s) cosmogenic isotopes.
Since \ber\ ($\tau=76.9$d, $Q$--value$=$0.9\,MeV) and \cele\ ($\tau=29.4$min, $Q$--value$=$2.0\,MeV) are below the energy threshold,  we focus on \cten\ and \bele.  

Taking into account the energy response of Borexino, the fraction of the \cten\ energy spectrum  above 3\,MeV  is 1.2\%.  When \cten\ is produced in association with a neutron, \cten\ candidates are tagged  by the three-fold coincidence with the parent muon and subsequent neutron capture in the scintillator~\cite{Gal05}.  The efficiency of the Borexino electronics in detecting at least one neutron soon after a muon has been estimated to be 94\% by two parallel (1-channel and 8-channel) DAQ systems that digitize data for 2\,ms after every OD trigger at 500\,MHz.  The rate of muons associated with at least 1 neutron, measured by the Borexino electronics, is  $\sim$67\,cpd.  Thus, to reject \cten\ from the analysis we exclude all data  within a 120\,s window after a $\mu$+$n$ coincidence and within a 0.8\,m distance from the neutron capture point.  The efficiency of this cut is 0.74$\pm$0.11, for a 0.16\% dead time.  A time profile analysis of  events tagged by this veto above 2.0\,MeV returns a characteristic time of 30$\pm$4\,s, consistent with the lifetime of \cten, and a total \cten\ rate of
(0.50$\pm$0.13)\,cpd/100\,t, in production channels with neutron emission.  Thus, the residual \cten\ contamination from neutron-producing channels above 3\,MeV is \powert{-3}{\mbox{(6.0$\pm$0.2)}}\,cpd/100~t.

The dominant  neutron-less \cten\ production reaction is   $^{12}$C($p$,$t$)$^{10}$C.  We extrapolated its rate by scaling the $^{12}$C($p$,$d$)$^{11}$C production rate \cite{Gal05,Gal052}, by the ratio between the $^{12}$C($p$,$d$)$^{11}$C and $^{12}$C($p$,$t$)$^{10}$C cross sections measured in Ref.~\cite{Yas77}.  The $^{12}$C($p$,$t$)$^{10}$C rate is \powert{-3}{\mbox{(0.6$\pm$1.8)}}\,cpd/100~t.  The residual background above 3\,MeV from \cten\ is \powert{-3}{2.2}\,cpd/100~t.

The overall \cten\ rate above 3 MeV, (6.6$\pm$1.8) $\times$10$^{-3}$\,cpd/100~t, agrees with the expected one,  quoted in Table~\ref{tab:cosmogenic}.

\subsubsection{\bele\ estimation}
\begin{figure}[t]
\includegraphics[width=\columnwidth]{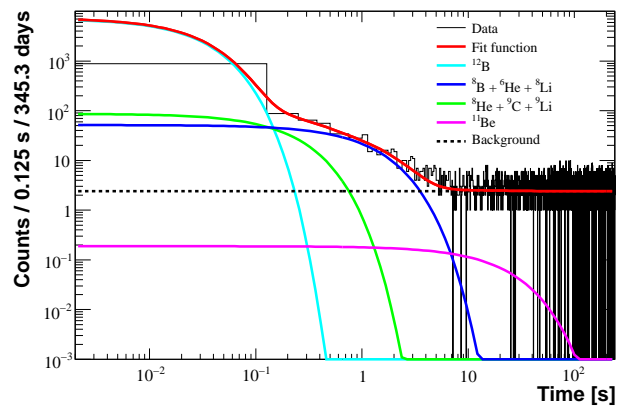}
\caption{Time profile of events  with energy $>$ 3 MeV within 240~s after a muon and within 2~m from its track  in the entire Borexino active volume (278~t). The distribution has been fit to the three decay exponentials as in Figure \ref{fig:TimeAfterMuon}, plus the $^{11}$Be component, with fixed mean-lives.  }
\label{fig:be11}
\end{figure}

Figure  \ref{fig:be11} shows the time profile of events within 240~s after a muon and within  a 2~m distance from its track  in the entire Borexino active volume (278~t). The efficiency of the distance cut  is assumed to be the same as the one measured for cosmogenic \btwe\ (84\%) by performing fits to the time distribution of events after a muon before and after  the track cut. The  measured \bele\ rate above 3~MeV is \powert{-2}{\mbox{(3.6$\pm$3.5)}}\,cpd/100\,t, consistent with the  \powert{-2}{\mbox{(3.2$\pm$0.6)}}\,cpd/100\,t  rate extrapolated from the KamLAND measurements~\cite{Abe09}. Since all measured rates deviated less than 18\% from the extrapolated value, we adopt the latter as the residual rate of \bele\  in our sample.

\subsection{Radioactive Background Rejection}

\begin{table}[!b]
\begin{center}\begin{tabular}{lcc}
\hline\hline
Cut				&Counts			&Counts \\
					
				&3.0--16.3~MeV	&5.0--16.3~MeV \\
\hline
All counts			&1932181	&1824858 \\
\textit{Muon and neutron cuts}	&6552		&2679 \\
\textit{FV cut}			&1329		&970 \\
\textit{Cosmogenic cut}		&131		&55 \\
\cten\ removal			&128		&55 \\
\bis\ removal			&119		&55 \\
\tal\ subtraction		&90$\pm$13	&55$\pm$7 \\
$^{11}$Be subtraction	        &79$\pm$13	&47$\pm$8 \\
Residual subtraction		&75$\pm$13	&46$\pm$8 \\
\hline
Final sample    		&75$\pm$13	&46$\pm$8 \\
\hline
BPS09(GS98) \bor\ $\nu$		&86$\pm$10	&43$\pm$6 \\
BPS09(AGS05) \bor\ $\nu$        &73$\pm$7	&36$\pm$4 \\
\hline\hline
\end{tabular}\end{center}
\caption{Effect of the sequence of cuts on the observed counts.  The cosmogenic cut introduces a reduction of the detector live-time of~29.2\%.  The resulting effective live-time is  345.3\,d.  The analysis is done with 100\,t fiducial mass target, after the FV cut.  The expected \bor\ counts are calculated from current best parameters for the MSW-LMA~\cite{Fog08} and Standard Solar Models, BPS09(GS98) and BPS09(AGS05)~\cite{BS07,Pen08,Ser09}}
\label{tab:cuts}
\end{table}

\subsubsection{External background}

\begin{figure}
\includegraphics[width=\columnwidth]{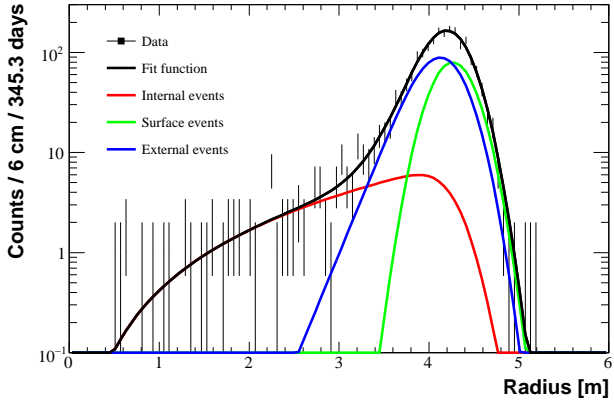}
\caption{Fit of the radial distribution for events with E$>$3~MeV. The red line represents the uniformly distributed event component in the active mass, the green line the surface contamination, and the blue line is external background.}
\label{fig:radfit}
\end{figure}

The 3\,MeV energy threshold is set  by the 2.614\,MeV $\gamma$-rays from the $\beta$-decay of \tal, due to radioactive contamination in the PMTs and in the SSS.  
Above 3\,MeV, the sources of radioactive background include the radioactive decays of residual \bis\ (\ura\ chain, $Q$=3.272\,MeV) and \tal\ (\tho\ chain, $Q$=5.001\,MeV) in the liquid scintillator.  The fiducial volume cut is very effective against the \tal\ and \bis\ background due to  \radon\ and \thoron\ emanated from the nylon vessel, as well as residual external $\gamma$-ray background.  In Figure~\ref{fig:radfit} the radial distribution of all scintillation events above 3\,MeV   has been fit to a model which takes into account the three sources of backgrounds:  a uniform distribution in the detector for internal events, a delta-function centered on the vessel radius for the point-like radioactive background in the nylon, and an exponential for external $\gamma$-ray background. All the three components are convoluted with the detector response function.  From this radial analysis we conclude that within the fiducial volume there is a small contribution of events  from surface contamination and the exterior of \powert{-3}{(6.4$\pm$0.2)}\,cpd/100~t (\powert{-6}{(3$\pm$11)}\,cpd/100~t) for events above 3\,MeV (5\,MeV).

\subsubsection{\bifo\ contamination}

The suppression of  \bifo\ from \ura\ contamination in the FV relies on the  \Bipo\ delayed coincidence ($\tau$=237\,$\mu$s).
We look for coincidences in time between 20\,$\mu$s and 1.4\,ms ($\epsilon$ = 0.91) with a spatial separation $<$1.5\,m ($\epsilon$ = 1) and Gatti parameter, a pulse shape discrimination estimator introduced in Ref.~\cite{BXD08,Bac08}, larger than -0.008 ($\epsilon$ = 1) within the fiducial volume throughout the entire data set. The \pofo\ $\alpha$--decays are selected in the 0.3--1.2\,MeV ($\epsilon$ = 1) energy range. The remaining contribution of \bifo\ to the $\nu$-$e$ scattering sample is  negligible (1.1$\pm$0.4)$\times$10$^{-4}$\,cpd/100\,t.

\subsubsection{\tal\ contamination}

Amongst the daughters of  \tho\  naturally present in the scintillator,    \tal\ decays are the only ones   which contribute background  above 3 MeV. The parent of \tal\ is \bitwe\  $\alpha$-decays into \tal\ with a branching ratio of 36\% and a lifetime of $\tau$ = 4.47~min. In the second channel with branching ratio  64\%,  \bitwe\ $\beta$-decays into \potwe\ with a lifetime of $\tau=$ 431~ns. We estimate the \tal\ rate from the fast \bipo\ coincidences. \bipo\ events are selected in a time window between 400 and 1300 ns, with an efficiency of 0.35, and requiring a maximum spatial distance between the two events of 1 m ($\epsilon$= 1). \bitwe\ and \potwe\ are selected in [20--1200] p.e. ($\epsilon$ = 1) and [420--580] p.e. ($\epsilon=$ 0.93)  energy regions, respectively. The \potwe--$\alpha$ quenched energy is estimated from the \pote\ and \pofo\ peaks,  optimal signatures for the $\alpha$-quenching calibration. We found 21 \bipo\ coincidences in the entire data set, within the FV. Accounting for the efficiency of the selection cuts and the branching ratios of the \bitwe\ decays, this corresponds to a \tal\ contamination in our neutrino sample of 29$\pm$7 events, or a  (8.4$\pm$2.0)$\times$10$^{-2}$ cpd/100\,t rate.

\begin{figure}[]
\includegraphics[width=0.5\textwidth]{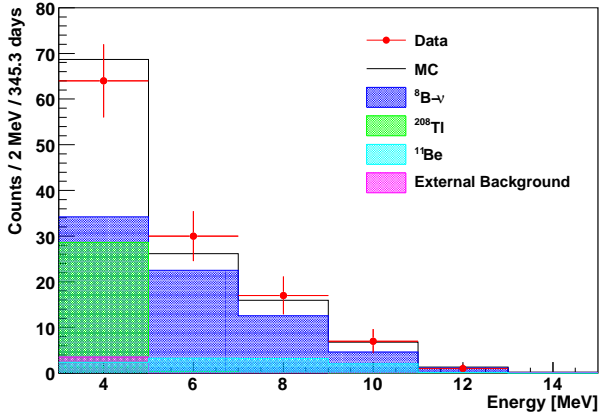}
\caption{Comparison of the final spectrum  after data  selection (red dots) to Monte Carlo simulations (black line).  The expected electron recoil spectrum from to oscillated \bor\ $\nu$ interactions (filled blue histogram),   $^{208}$Tl (green), $^{11}$Be (cyan) and external background (violet), are equal to the measured values  in Table \ref{tab:residual}.} 
\label{fig:b8spectrum}
\end{figure}

A summary of the analysis sequence described above is shown in Table~\ref{tab:cuts}. The energy spectrum of the final sample, compared with simulated spectra of \bor\ neutrinos and of each residual background component listed in Table~\ref{tab:residual}, is  shown in Figure~\ref{fig:b8spectrum}.

\begin{table}[!b]
\begin{center}\begin{tabular}{lll}
\hline\hline
Background &Rate [10$^{-4}$cpd/100 t]&	\\
           & $>$3~MeV& $>$5~MeV	\\
\hline
\textit{Muons}	         &4.5$\pm$0.9&3.5$\pm$0.8\\
\textit{Neutrons}        &0.86$\pm$0.01 & 0\\
\textit{External background} &64$\pm$2 &0.03$\pm$0.11\\
\textit{Fast cosmogenic} &17$\pm$2 &13$\pm$2\\
\cten\ 			 &22$\pm$2  & 0\\
\bis\ 			 &1.1$\pm$0.4 & 0\\
\tal\ 			 &840$\pm$20  & 0\\
\bele\ 			 &320$\pm$60 & 233$\pm$44\\
\hline
\textit{Total}		 &1270$\pm$63 &250$\pm$44\\
\hline\hline
\end{tabular}\end{center}
\caption{Residual rates of background components after the data selection cuts above 3 and 5 MeV.}
\label{tab:residual}
\end{table}

\section{Neutrino interaction rates and electron scattering spectrum}
\label{sec:neutrinos}
The mean value for \bor\ neutrinos in the sample above 3~MeV (5~MeV) is 75$\pm$13 (46$\pm$8) counts. 

The dominant sources of systematic errors are the determinations of the energy threshold and of the fiducial mass, both already discussed in the previous sections. The first  introduces a systematic uncertainty of  +3.6\% -3.2\% (+6.1\% -4.8\% above 5\,MeV).   The second systematic source is responsible for a $\pm$3.8\%  uncertainty in the \bor\ neutrino rate.  A secondary source of systematics, related to the effect of the energy resolution on the threshold cuts, has been studied on a simulated \bor\ neutrino spectrum and is responsible for a   systematic uncertainty of  +0.0\% -2.5\% (+0.0\% -3.0\% above 5\,MeV). 

The total systematic errors are shown in Table~\ref{tab:syst}.

\begin{figure}[]
\includegraphics[width=0.5\textwidth]{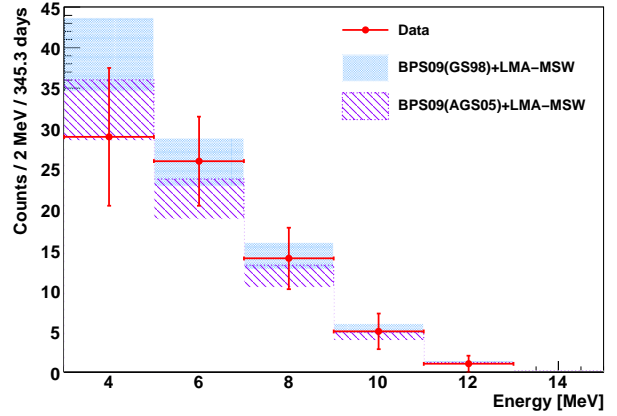}
\caption{Comparison of the final spectrum  after data  selection and background subtraction (red dots) to Monte Carlo simulations (blue) of oscillated \bor\ $\nu$ interactions, with amplitude from the Standard Solar Models BPS09(GS98) (high metallicity) and BPS09(AGS05) (low metallicity), and from the MSW-LMA neutrino oscillation model.} 
\label{fig:b8fromModel}
\end{figure}

The resulting count rate with E$>$3\,MeV  is:
$$0.217\pm 0.038 (stat)^{+0.008}_{-0.008}(syst)~cpd/100\,t$$ 
and with  E$>$5\,MeV:
$$0.134\pm 0.022(stat)^{+0.008}_{-0.007}(syst)~cpd/100\,t.$$
The final energy spectrum after all cuts and residual background is shown in Figure~\ref{fig:b8fromModel}. It is in agreement with the scenario which combines the high metallicity Standard Solar Model, called BPS09(GS98) \cite{Ser09},  and the prediction of the MSW-LMA solution.

\begin{table}
\begin{center}
\begin{tabular}{lcccc}
\hline\hline
Source					&E$>$3~MeV	& &E$>$5~MeV &  \\
 					&$\sigma_+$	&$\sigma_-$ &$\sigma_+$	&$\sigma_-$\\
\hline
Energy threshold			& 3.6\%  & 3.2\% & 6.1\% & 4.8\%\\
Fiducial mass				& 3.8\%  & 3.8\% & 3.8\% & 3.8\%\\
Energy resolution			& 0.0\%  & 2.5\% & 0.0\% & 3.0\%\\
\hline
Total                                  & 5.2\%   & 5.6\% & 7.2\% & 6.8\%\\
\hline\hline
\end{tabular}
\caption{Systematic errors.}
\label{tab:syst}
\end{center}
\end{table}

\section{Solar \bor\ neutrino flux and neutrino oscillation parameters}
\label{sec:implications}

The equivalent unoscillated \bor\ neutrino flux,  derived from the electron scattering rate above 5~MeV (Table~\ref{tab:comparison}), is (2.7$\pm$0.4$_{\rm stat}$$\pm$0.2$_{\rm syst}$)$\times$10$^6$~cm$^{-2}$s$^{-1}$, in good agreement with the SuperKamiokaNDE-I and SNO~D$_2$O measurements with the same threshold, as reported in Table \ref{tab:fluxes}.  The corresponding value above 3\,MeV, is (2.4$\pm$0.4$_{\rm stat}$$\pm$0.1$_{\rm syst}$)$\times$10$^6$~cm$^{-2}$s$^{-1}$.  The expected value for the case of no neutrino oscillations, including the theoretical uncertainty  on the \bor\ flux from the Standard Solar Model~\cite{BS07,Pen08,Ser09}, is  (5.88$\pm$0.65)$\times$10$^6$~cm$^{-2}$s$^{-1}$ and, therefore, solar $\nu_e$ disappearance is confirmed at 4.2$\sigma$.

To define the neutrino electron survival probability \oPee\ averaged in the energy range of interest, we define the  measured recoiled electron rate $R$, through the convolution:

\begin{equation}
R = \int_{T_e>T_0} dT_e \int_{{ - \infty}}^{{ + \infty}} dt \left(\frac{dr}{dt}(t) \cdot g(t-T_e)\right) 
\label{eq:meanpee}
\end{equation}
between  the detector energy response $g$, assumed gaussian, with a resolution depending on the energy, 
and differential rate:
 
\begin{equation}
\frac{dr}{dT_e}(T_e) = \int_{E_{\nu}>E_{0}} dE_{\nu} \left( \overline P_{ee}\cdot\Psi_{e}   + (1 - \overline P_{ee}) \cdot \Psi_{\mu-\tau} \right)
\label{eq:meanpee}
\end{equation}

\noindent with:

\begin{equation}
\Psi_{e} =\frac{d\sigma_e}{dT_e}(E_{\nu},T_e) \cdot N_e \cdot \frac{d\Phi_e}{dE_{\nu}} (E_{\nu}),
\label{eq:meanpee}
\end{equation}

\noindent and:

\begin{equation}
\Psi_{\mu-\tau} = \frac{d\sigma_{\mbox{$\mu$-$\tau$}}}{dT_e}(E_{\nu},T_e)\cdot N_e \cdot \frac{d\Phi_e}{dE_{\nu}} (E_{\nu}).
\label{eq:meanpee}
\end{equation}

\noindent where $T_e$ and $E_{\nu}$ are the electron and neutrino energies, and $\sigma_{x}$ ($x$=$e$,$\mu$-$\tau$) are the cross sections for elastic scattering for different flavors.  $T_0$=3\,MeV is the energy threshold for scattered electrons, corresponding to a minimum neutrino energy of $E_0$=3.2\,MeV, $N_e$ is the number of target electrons, and $d\Phi_e/dE_{\nu}$ is the differential \bor\ solar neutrino flux \cite{Bah96}.

\begin{table}
\begin{center}\begin{tabular}{lcc}
\hline\hline
		                                                &3.0--16.3\,MeV                 &5.0--16.3\,MeV \\
\hline
Rate [cpd/100 t]	                                        & 0.22$\pm$0.04$\pm$0.01        & 0.13$\pm$0.02$\pm$0.01 \\
$\Phi^{\rm ES}_{\rm exp}$ [10$^6$~cm$^{-2}$s$^{-1}$]            & 2.4$\pm$0.4$\pm$0.1	        &2.7$\pm$0.4$\pm$0.2 \\
\textit{$\Phi^{\rm ES}_{\rm exp} / \Phi^{\rm ES}_{\rm th}$}	& 0.88$\pm$0.19			&1.08$\pm$0.23 \\
\hline\hline
\end{tabular}
\end{center}
\caption{Measured event rates in Borexino and comparison with the expected theoretical flux in the BPS09(GS98) MSW-LMA scenario \cite{MSW}.}
\label{tab:comparison}
\end{table}

Using the above equation, we obtain \mbox{\oPee}=0.29$\pm$0.10 at the mean energy of 8.9\,MeV for \bor\ neutrinos.   Borexino is the first experiment to detect in real time, and  in the same target, neutrinos in the low energy, vacuum dominated- and in the high energy, matter enhanced-regions.   Borexino already reported a survival probability of 0.56$\pm$0.10 for \ber\ neutrinos, at the energy of 0.862\,MeV~\cite{BX08}. The distance between the two survival probabilities is 1.9 $\sigma$.

\begin{figure}
\includegraphics[width=\columnwidth]{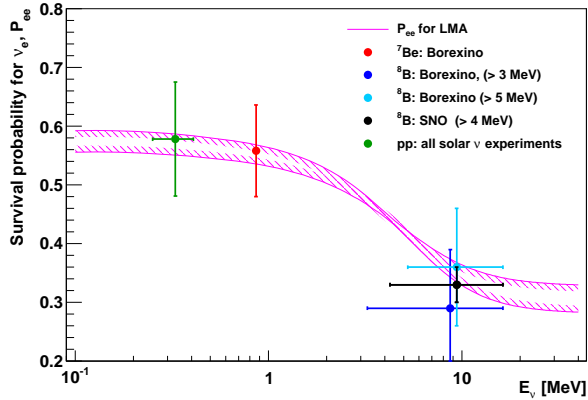}
\caption{Electron neutrino survival probability as function of the neutrino energy, evaluated for the \bor\ neutrino source assuming the BPS09(GS98) Standard Solar Model~\cite{BS07,Pen08,Ser09} and the oscillation parameters from the MSW-LMA solution $\Delta$m$^2$=7.69$\times$10$^{-5}$~eV$^2$ and $\tan^2{\theta}$=0.45~\cite{Fog08}).  Dots represent the Borexino results from \ber\ and \bor\ measurements. The error bars include also the theoretical uncertainty of the expected flux from the  Standard Solar Model BPS09(GS98). }
\label{fig:pee}
\end{figure}

Future precision measurements of \ber\ and \bor\ (and, possibly, \pep) neutrinos in Borexino could provide an even more stringent test of the difference in \Pee\ for low- and high-energy neutrinos predicted by the MSW-LMA theory: assuming other 4 years of data taking, and to reduce the overall uncertainty on \ber\ neutrino rate at 5\%, the distance between the two survival probabilities can be improved at 3 $\sigma$.

\begin{table}
\begin{center}
\begin{tabular}{lcc}
\hline\hline
					&Threshold	&$\Phi^{\rm ES}_{\rm ^8B}$ \\
					&[MeV]		&[10$^6$~cm$^{-2}$ s$^{-1}$] \\
\hline
SuperKamiokaNDE~I~\cite{SKI05}	        &5.0            &2.35$\pm$0.02$\pm$0.08 \\
SuperKamiokaNDE~II~\cite{SKII08}	&7.0            &2.38$\pm$0.05$^{+0.16}_{-0.15}$ \\
SNO~D$_2$O~\cite{SNO07}		        &5.0		&2.39$^{+0.24}_{-0.23}$ $^{+0.12}_{-0.12}$ \\
SNO~Salt~Phase~\cite{SNO05}	        &5.5		&2.35$\pm$0.22$\pm$0.15 \\
SNO~Prop.~Counter~\cite{SNO08}	        &6.0	        &1.77$^{+0.24}_{-0.21}$$^{+0.09}_{-0.10}$ \\
Borexino				&3.0            &2.4$\pm$0.4$\pm$0.1 \\
Borexino				&5.0            &2.7$\pm$0.4$\pm$0.2 \\
\hline\hline
\end{tabular}
\caption{Results on \bor\ solar neutrino flux from elastic scattering, normalized under the assumption of the no-oscillation scenario reported by SuperKamiokaNDE, SNO, and Borexino.}
\label{tab:fluxes}
\end{center}
\end{table}

We are grateful to F.~Vissani and F.~Villante for useful discussions and comments. The Borexino program was made possible by funding from INFN (Italy), NSF (U.S., PHY-0802646,  PHY-0802114, PHY-0902140), BMBF, DFG and MPG (Germany), Rosnauka (Russia), MNiSW (Poland). We acknowledge the generous support of the Laboratori Nazionali del Gran Sasso (LNGS). This work was partially supported by PRIN 2007 protocol 2007JR4STW.


\end{document}